\documentclass[journal=jctc,manuscript=article,layout=twocolumn]{achemso}

\usepackage{achemso}
\usepackage{graphicx}
\usepackage{dcolumn}
\usepackage{bm}
\usepackage{amsmath}
\usepackage[capitalise]{cleveref}
\SectionNumbersOn

\title{Toward Orbital-Free Density Functional Theory with Small Datasets and Deep Learning}

\author{Kevin Ryczko}
 \email{kevin.ryczko@uottawa.ca}
 \affiliation{Department of Physics, University of Ottawa, Ottawa, Ontario, Canada}
 \affiliation{1QB Information Technologies (1QBit), Vancouver, Bristish Columbia, Canada}
 \affiliation{Vector Institute for Artificial Intelligence, Toronto, Ontario, Canada}

\author{Sebastian J. Wetzel}
\affiliation{Perimeter Institute for Theoretical Physics, Waterloo, Ontario, Canada}

\author{Roger G. Melko}
\affiliation{Perimeter Institute for Theoretical Physics, Waterloo, Ontario, Canada}
\affiliation{Department of Physics and Astronomy, University of Waterloo, Waterloo, Ontario, Canada}

\author{Isaac Tamblyn}
\email{isaac.tamblyn@uottawa.ca}
\affiliation{Department of Physics, University of Ottawa, Ottawa, Ontario, Canada}
\affiliation{Vector Institute for Artificial Intelligence, Toronto, Ontario, Canada}

\date{\today}
\begin{document}

\begin{abstract}
We use voxel deep neural networks to predict energy densities and functional derivatives of electron kinetic energies for the Thomas-Fermi model and Kohn-Sham density functional theory calculations. We show that the ground-state electron density can be found via direct minimization for a graphene lattice without any projection scheme using a voxel deep neural network trained with the Thomas-Fermi model. Additionally, we predict the kinetic energy of a graphene lattice within chemical accuracy after training from only 2 Kohn-Sham density functional theory (DFT) calculations. We identify an important sampling issue inherent in Kohn-Sham DFT calculations and propose future work to rectify this problem. Furthermore, we demonstrate an alternative, functional derivative-free, Monte Carlo based orbital free density functional theory algorithm to calculate an accurate 2-electron density in a double inverted Gaussian potential with a machine-learned kinetic energy functional.
\end{abstract}

\section{Introduction}\label{intro}

Kohn-Sham density-functional theory \cite{kohn1965self} (KS-DFT) and Orbital-Free (OF) DFT \cite{wang2002orbital, ligneres2005introduction} are two electronic structure methodologies to calculate properties of matter. In OF-DFT, all energy functionals depend solely on the electron density, whereas in KS-DFT, energy functionals depend on both the electron density and the set of Kohn-Sham orbitals. The explicit dependence on the electron density in OF-DFT allows for favourable, $\mathcal{O}(N)$, computational scaling, enabling one to study large systems \cite{hung2009accurate} (where $N$ is the number of electrons). Conversely, The computational scaling of KS-DFT, $\mathcal{O}(N^3)$, is less favourable due to the computation of a set of orbitals, rather than the electron density alone. However, the main advantage of KS-DFT implementations is that the kinetic energy is calculated via a single-particle quantum mechanical operator, leading to a more accurate approximation of the true kinetic energy functional (KEF). In OF-DFT, the kinetic energy is written as a classical, approximate functional of the electron density. The lack of knowledge of a quantum mechanical KEF reduces the accuracy and applicability of OF-DFT. \\

Thomas and Fermi (TF) both proposed an analytic KEF assuming a free electron gas \cite{thomas1927calculation, fermi1927atti}. They were followed by the Thomas-Fermi-Dirac-von Weizs\"acker and $X\alpha$ models \cite{gombas1956handbuch, march1957thomas, lieb1982erratum, dunlap1979first} to address the failures of the TF model for atoms and molecules. Hohenberg and Kohn \cite{hohenberg1964inhomogeneous} proved the existence of a KEF that depends explicitly on the electron density of interacting electrons, but never gave its exact functional form. Subsequently, Kohn and Sham \cite{kohn1965self} introduced a non-interacting, orbital-dependant KEF. This non-interacting functional is routinely used in all KS-DFT calculations. \\

More recently, machine learning models have been used as energy functionals \cite{snyder2012finding, brockherde2017bypassing, mayer2020machine, nagai2020completing, kalita2021learning, schleder2019dft}. Specifically, in Refs. \cite{snyder2012finding, mayer2020machine} machine-learned, one-dimensional KEFs were constructed using kernel ridge regression and convolutional neural networks (CNNs). In Ref. \cite{snyder2012finding}, the authors argued that the error of a functional derivative of a machine-learned KEF (FD-KEF) was too large to be used in a direct minimization calculation. They reduced this error by projecting the functional derivative of the total energy onto a subspace found with principal component analysis. Following this report, Ref. \cite{mayer2020machine} included the FD-KEF in a loss function to improve the predictions from the machine learning models. This improved loss function reduced the prediction error of the FD-KEF but did not eliminate it entirely. An additional projection method using a sinusoidal basis was introduced and utilized to minimize the error. The use of a sinusoidal basis eliminated the computational overhead of performing principal component analysis on the training set densities. \\

In addition to KEFs, machine learning models have been used as exchange-correlation functionals \cite{zhou2019toward, ryabov2020neural, lyon2020fundamentals}. In Refs. \cite{zhou2019toward, ryabov2020neural}, ``slices" of the density, rather than the entire scalar field, were used as input to neural networks. It was shown that machine-learned exchange-correlation functionals could be used for a model system with a simple harmonic oscillator potential, several molecules, and a unit cell of Si, demonstrating the transferability of this methodology. Additionally, the approach drastically reduced the number of calculations needed to generate a training set.\\ 

\begin{figure}
    \centering
    \includegraphics[width=\linewidth]{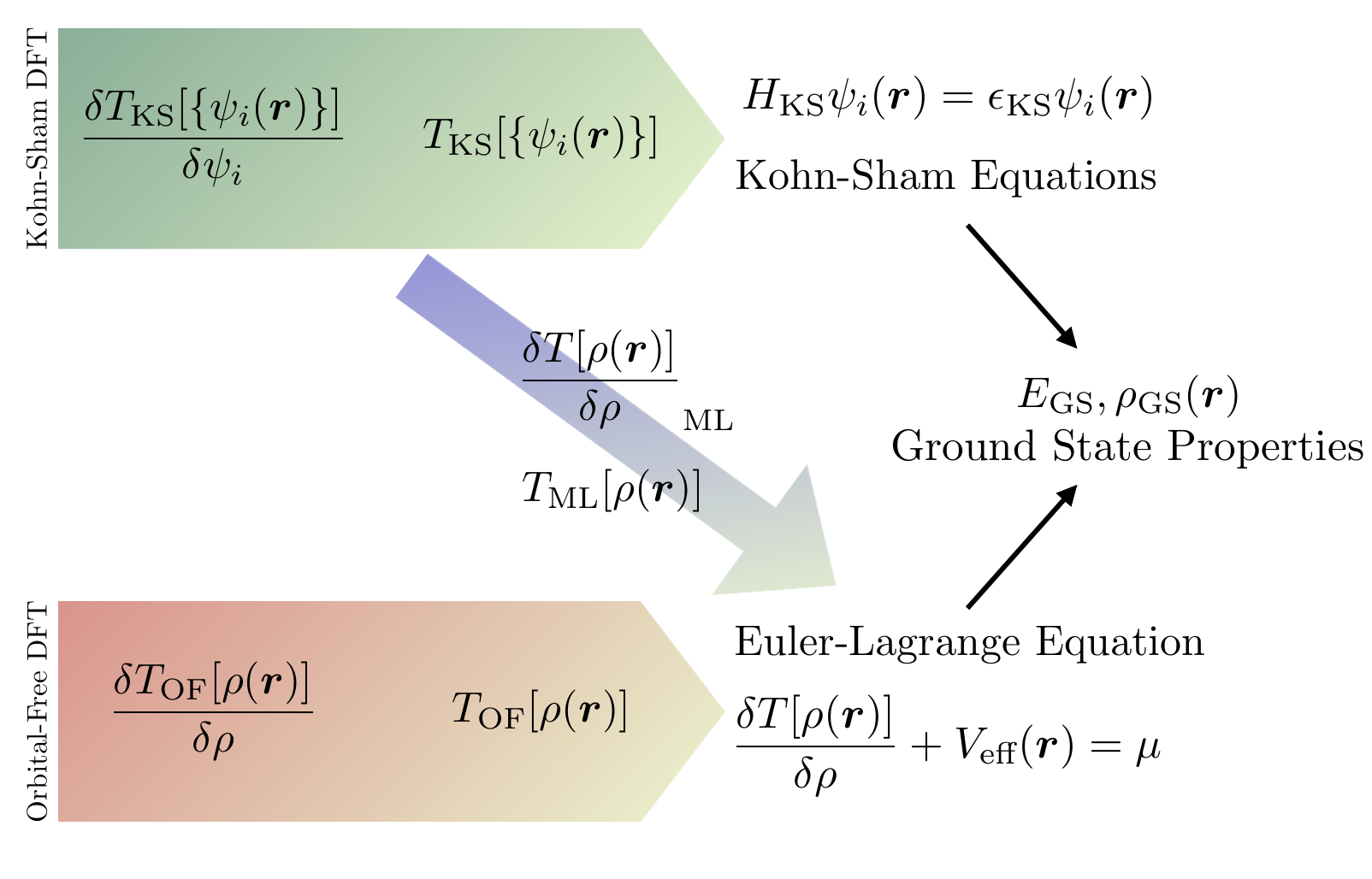}
    \caption{Our machine learning architecture, similar to Refs. \cite{snyder2012finding, mayer2020machine}, makes a connection between Kohn-Sham density functional theory and orbital-free density functional theory. The model allows for the construction of Kohn-Sham kinetic energy functionals that explicitly depend on the electron density and, therefore, direct insertion into orbital-free density functional theory. See \cref{methods} for more information about the equations. }
    \label{fig_0}
\end{figure}

\begin{figure*}[ht]
    \centering
    \includegraphics[width=\linewidth]{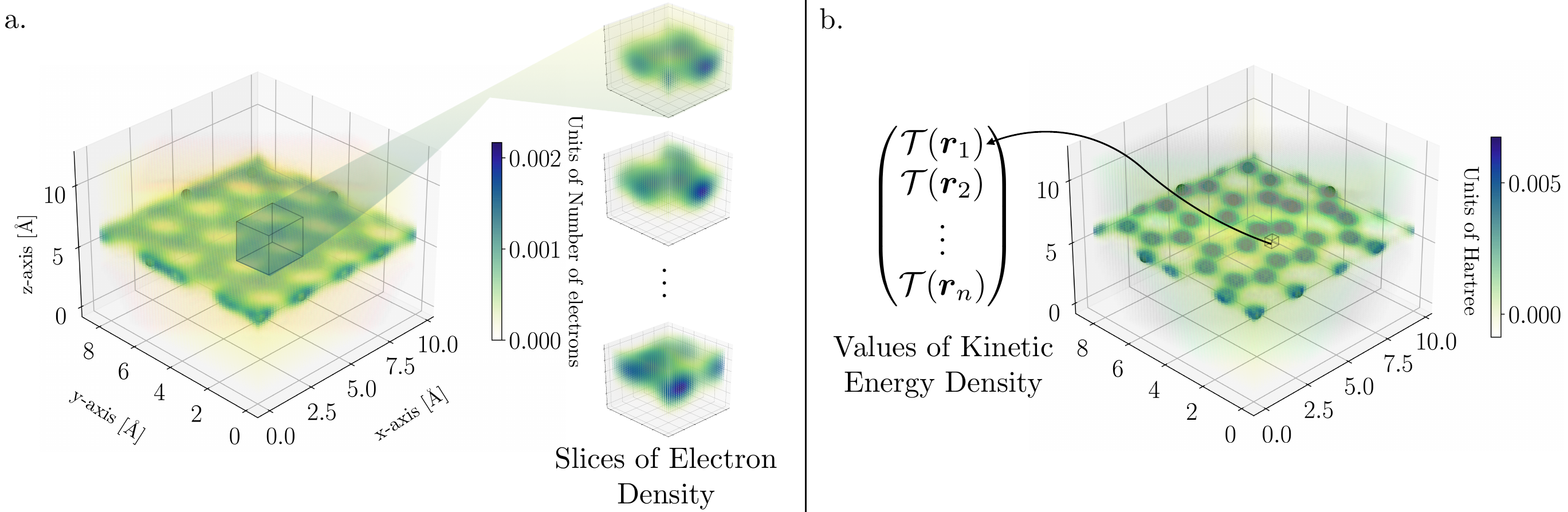}
    \caption{A visual representation of voxel deep neural networks. (a) An example electron density for a 32 atom graphene lattice supercell. The highlighted region in the electron density is a slice of the electron density centered at a particular pixel.  (b) The kinetic energy density for the same 32 atom graphene lattice. The voxel deep neural network learns the mapping between the slice of electron density to the voxel of kinetic energy density.}
    \label{fig:methods}
\end{figure*}

We build on previous work which computed KEFs for one-dimensional systems and compute KEFs, FD-KEFs, electron densities, and energies in \textit{three dimensions for a realistic system}: pristine graphene lattices. We also \textit{eliminate the need for large datasets}. Namely, we use slices of the electron density as input to deep neural networks (DNNs), where the output is also a slice of the kinetic energy density (KED). Desired quantities are subsequently found via integration over the supercell. We call this methodology voxel DNNs (VDNNs). In Section \ref{methods}, we outline the basic electronic structure, training data generation, and machine learning methodologies used. In Section \ref{results}, we outline the results of VDNNs in practice. We first investigate the Thomas-Fermi model with VDNNs as a proof of principle. The Thomas-Fermi model is simple and both the kinetic energy and its functional derivative with respect to the electron density are analytically known for all densities. Afterwards, we apply VDNNs to KS-DFT. Using VDNNs allows one to have a Kohn-Sham kinetic energy functional that explicitly depends on the electron density and enables one to insert the energy functional into OF-DFT (\cref{fig_0}). Lastly, we show an alternative potential of our method with a demonstration of a functional derivative-free Monte Carlo (MC) based optimization for a toy, 1D model system. Direct minimization techniques have been applied in KS-DFT calculations \cite{yang2006constrained, weber2008direct} which avoids the self-consistent procedure, but direct minimization in OF-DFT still requires functional derivatives. Our MC based optimization eliminates the need of a functional derivative altogether. We conclude and propose future directions based on our results in \cref{conclusion}. 

\section{Methods}
\label{methods}
In this work, we use VDNNs to calculate KEDs ($\mathcal{T}$) and FD-KEFs ($\mathcal{F}$) of graphene lattices using OF-DFT with the Thomas-Fermi model and using KS-DFT (LDA and GGA). As discussed above, the Thomas-Fermi model serves as a preliminary experiment due to its simplicity and KS-DFT serves as a realistic use case. We therefore first test our methodology with the Thomas-Fermi model before moving on to KS-DFT. In OF-DFT, the total energy functional is written in real space as
\begin{eqnarray}
\label{total_energy}
    E[\rho(\bm{r})] = T[\rho(\bm{r})] + E_{\text{Hartree}}[\rho(\bm{r})]\nonumber\\
	 + E_{\text{ion}}[\rho(\bm{r})] + E_{\text{xc}}[\rho(\bm{r})]
\end{eqnarray}
where $\rho(\bm{r})$ is the electron density and the terms in order are kinetic, Hartree, external, and exchange-correlation energies. To find the ground state electron density, one searches for an electron density which minimizes the total energy expression under the constraint that the number of electrons, $N_e$, is fixed. This yields the Lagrangian
\begin{equation}
    \mathcal{L}[\rho(\bm{r})]=E[\rho(\bm{r})] -\mu \left(\int_{\Omega}d\bm{r}~\rho(\bm{r}) - N_e\right)
\end{equation}
where $\mu$ is a Lagrange multiplier and $\Omega$ is the volume of the supercell. Applying a functional derivative of the Lagrangian with respect to the electron density yields the Euler-Lagrange equation
\begin{equation}
\label{euler-lagrange}
    \mathcal{F}(\bm{r}) + V_{\text{eff}}(\bm{r}) = \mu,
\end{equation}
where
\begin{equation}
    V_{\text{eff}}(\bm{r}) = V_{\text{Hartree}}(\bm{r}) + V_{\text{ion}}(\bm{r}) + V_{\text{xc}}(\bm{r}),
\end{equation}
and
\begin{equation}
    \mathcal{F}(\bm{r})=\frac{\delta T[\rho](\bm{r})}{\delta \rho(\bm{r})}.
\end{equation}
Using gradient descent, one can solve for the ground state electron density via direct minimization 
\begin{equation}
    \label{opt}
    \phi_{n + 1}(\bm{r}) = \phi_n(\bm{r}) - 2\alpha \phi_n(\bm{r})\left(\mathcal{F}(\bm{r}) + V_{\text{eff}}(\bm{r}) - \mu\right)_n
\end{equation}
where $\phi_n(\bm{r}) = \sqrt{\rho_n(\bm{r})}$, and $\alpha$ is a small parameter. The use of the square root of the density ensures that the electron density remains positive during the optimization. \\

Using the Thomas-Fermi model and the DFTpy code \cite{shao2020dftpy}, we performed 2 direct minimization calculations for 32-atom slabs of graphene where the atoms were perturbed from their equilibrium geometry. The perturbations were generated from a normal distribution with a standard deviation of 0.1 \AA. We used an energy cutoff of 45 Ha, the LDA exchange-correlation functional \cite{kohn1965self}, and norm-conserving pseudopotentials \cite{van2018pseudodojo}. Due to the free electron gas approximation used for the kinetic energy, we maintained this approximation in our exchange-correlation functional choice. We collected $\rho_{\text{TF}}$, $\mathcal{T}_{\text{TF}}$, and $\mathcal{F}_{\text{TF}}$ every 10 steps (values of $n$ in \cref{opt}) from one of the direct minimization calculations to be used as training data. This made for a total of 173 training configurations. The second calculation was used as independent test data.\\

In addition to OF-DFT calculations, we used KS-DFT to investigate 32-atom graphene slabs where the atoms were perturbed in the same way as described above. In KS-DFT, the electron density is written as
\begin{equation}
\label{rho}
    \rho_{\text{KS}}(\bm{r}) = 2\sum_{n}^{\text{occ}}\sum_{\bm{k}}w_{\bm{k}} \psi_{n, \bm{k}}^*(\bm{r})\psi_{n, \bm{k}}(\bm{r})
\end{equation}
and the KED is written as
\begin{equation}
\label{ke}
    \mathcal{T}_{\textsc{KS}}(\bm{r}) = -\sum_{n}^{\text{occ}}\sum_{\bm{k}}w_{\bm{k}} \psi_{n, \bm{k}}^*(\bm{r})\nabla^2\psi_{n, \bm{k}}(\bm{r}).
\end{equation}
In Equations \ref{rho} and \ref{ke}, $n$ is the band index, $\bm{k}$ is the k-point, $w_{\bm{k}}$ is the weighting associated with each k-point and $\psi$ is a Kohn-Sham orbital. In this work, we use finite differences to compute derivatives of the Kohn-Sham orbitals. To compute the Kohn-Sham orbitals we used Abinit \cite{Gonze2020} with an energy cutoff of 45 Ha, a $4\times4\times1$ k-point grid, the PBE exchange-correlation functional \cite{perdew1996generalized} and norm-conserving pseudopotentials \cite{van2018pseudodojo}. We justify this exchange-correlation choice based on its popularity in the literature. Here, we performed a total of 200 DFT calculations where 100 of the configurations were for training and 100 for kept aside for testing. To obtain $\mathcal{F}_{\text{KS}}$ for these calculations, we used \cref{euler-lagrange} where the potentials were evaluated using DFTpy \cite{shao2020dftpy}, and the chemical potentials were obtained from Abinit. It should be noted that \cref{euler-lagrange} can only be used to define $\mathcal{F}_{\text{KS}}$ when self-consistency has been reached \cite{liu2004functional}. \\

To train the VDNNs, we collected slices of $\rho$ (and $\nabla \rho$ for Kohn-Sham models) as inputs and slices of $\mathcal{T}$ and $\mathcal{F}$ as outputs. If $\widetilde{\mathcal{T}}$ and $\widetilde{\rho}$ are discretized forms of $\mathcal{T}$ and $\rho$ then a slice of $\rho$ with dimensions $(a, b, c)$ centred at pixels $(i,j,k)$ is written as $\widetilde{\rho}[i - a/2:i + a/2 + 1, j - b/2: j + b/2 + 1, k - c/2: k + c/2 + 1]$. The addition of 1 is due to the use of odd values of $a,b,c$. A slice of $\mathcal{T}$ with dimensions $(a', b', c')$ centred at pixels $(i, j, k)$ is similarly written as $\widetilde{\mathcal{T}}[i - a'/2:i + a'/2 + 1, j - b'/2: j + b'/2 + 1, k - c'/2: k + c'/2 + 1]$. We tested a variety of input sizes and used output sizes of (1,1,1). This corresponds to mapping electron density slices to values of $\mathcal{T}$, as shown in Figure \ref{fig:methods}. To avoid bias in training, we sample $\mathcal{T}$ or $\mathcal{F}$ such that a uniform distribution is produced given a target number of samples. The target number of samples was $1024^2$ unless stated otherwise. Of these, 99\% of them were used for training, and 1\% were used for validation. Testing was done on the 100 independent DFT calculations not seen during training. Inputs were standardized and normalized such that the range of values was $\in[-1,1]$ and outputs were normalized $\in[0,1]$. We used a modified version of the deep neural network (DNN) architecture used in Refs. \cite{ryczko2018convolutional, ryczko2019deep}, which had success in predicting various energies at the DFT level with different functionals. This included 2 non-reducing convolutional layers with 64 $3\times3\times3$ kernels, 4 non-reducing convolutional layers with 16 $3\times3\times3$ kernels, a reducing convolutional layer with 64 $3\times3\times3$ kernels, 4  non-reducing convolutional layers with 32 $3\times3\times3$ kernels, a fully connected layer with 1024 neurons, and a fully connected layer with 2 outputs. We use the ELU activation function throughout due to its improved performance compared to RELU with batch normalization \cite{klambauer2017self}. Since the inputs are scalar fields, a natural architectural choice is to use convolutional layers. They are designed to extract relevant features from images to make accurate predictions. The input dimensions are less than in Refs. \cite{ryczko2018convolutional, ryczko2019deep}, which is why the first 2 convolutional layers were changed to non-reducing layers. We note that this particular architecture choice is most likely not optimal, and one could obtain better results with another architecture choice. Models were trained for 500 epochs with learning rates of $10^{-5}$ and a batch size of 512. Production models were trained across 16 NVIDIA V100 GPUs with layer-wise adaptive rate scaling with clipping \cite{you2017scaling}. Training on large batch sizes leads to unfavourable results and Ref. \cite{you2017scaling} have shown that layer-wise adaptive rate scaling allows one to obtain similar results to lower batch training while reducing the training time. Inference for the densities can be trivially parallelized and does not suffer from any negative large-batch effects. It was done across 64 NVIDIA V100 GPUs. Our method does not require this GPU setup, but can make use of them when performing inference on large grids. Our multi-node, multi-GPU training code and our multi-node, multi-GPU inference code can be found here \cite{codes}.  \\

\section{Results}
\label{results}

\begin{figure}[t]
    \centering
    \includegraphics[width=\linewidth]{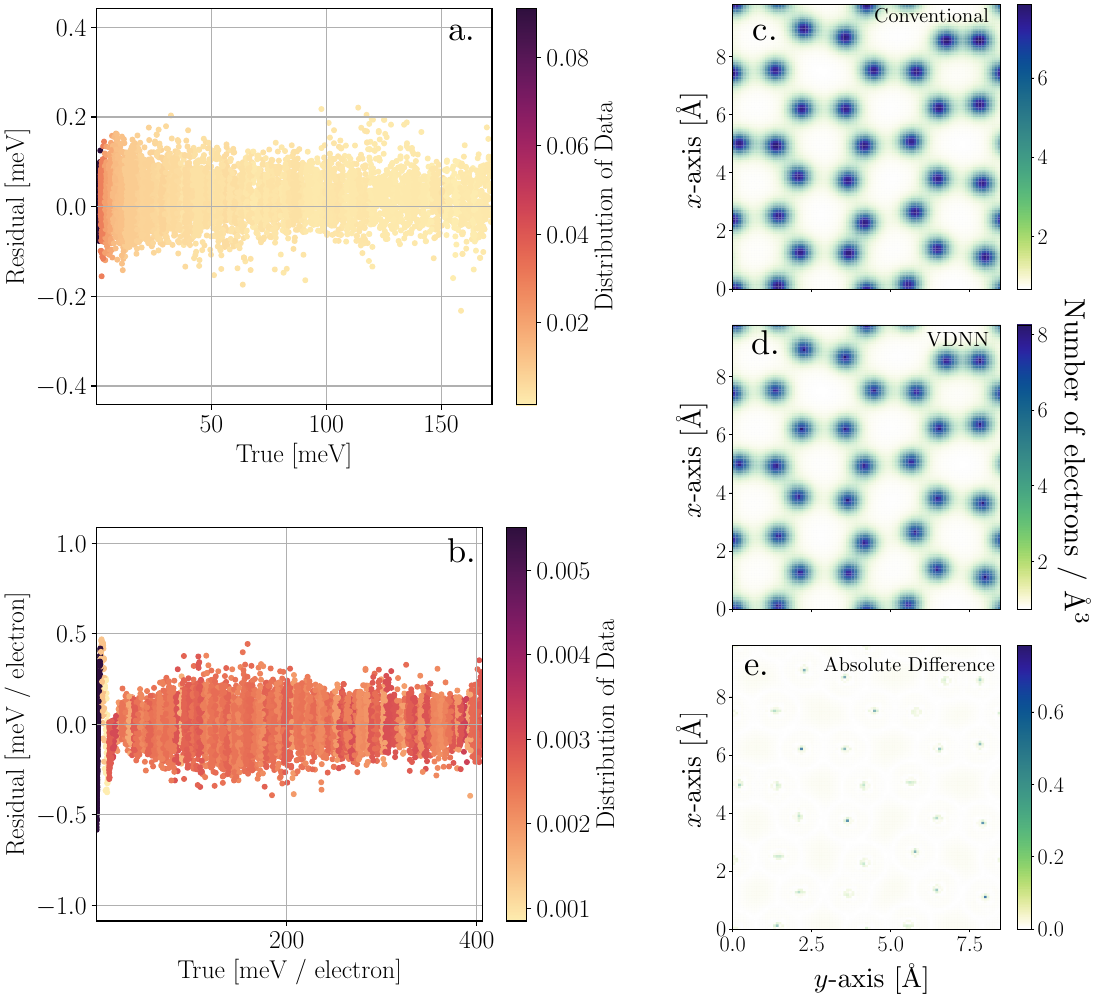}
    \caption{Thomas-Fermi model: Residual (true minus predicted) versus true for (a) $\mathcal{T}_{\text{TF}}$ and (b) $\mathcal{F}_{\text{TF}}$. (c) Thomas-Fermi electron density from a traditional direct minimization calculation and (d) electron density from a direct minimization calculation with a VDNN trained from a single OF-DFT calculation. (e) Absolute differences between the densities. VDNNs can be used in direct minimization calculations to find electron densities for the Thomas-Fermi model.}
    \label{fig:TF}
\end{figure}

We first discuss using VDNNs for the TF model. After training on $\mathcal{T}_{\text{TF}}$ and $\mathcal{F}_{\text{TF}}$ simultaneously, where $\mathcal{F}_{\text{TF}}$ was uniformly sampled, we study the accuracy of the model on the validation and testing data. Looking to \cref{fig:TF}a-b, we plot residuals for $\mathcal{T}_{\text{TF}}$ and $\mathcal{F}_{\text{TF}}$ for the validation set in units of meV and meV / electron, respectively. Density values have been multiplied by the volume such that direct integration over the numerical grid yields units of energy or energy / electron. From these plots, we can see that the residuals are a small fraction of their respective ranges. MAEs for $\mathcal{T}_{\text{TF}}$ and $\mathcal{F}_{\text{TF}}$ are 0.04 meV, and 0.08 meV / electron. RMSEs for $\mathcal{T}_{\text{TF}}$ and $\mathcal{F}_{\text{TF}}$ are 0.05 meV, and 0.11 meV / electron. The error for $\mathcal{F}_{\text{TF}}$ is larger than $\mathcal{T}_{\text{TF}}$. Part of this increase in error can be attributed to the increase in the range of values (a factor of 2.67 from $\mathcal{T}_{\text{TF}}$ to $\mathcal{F}_{\text{TF}}$), which contributes to 96\% of the increased error; the remaining increase in error is due to the VDNN.

We now use VDNNs to calculate an electron density and energy for the second, testing configuration via Equation \cref{euler-lagrange}. In past reports \cite{snyder2012finding, meyer2020machine}, it was declared unfeasible to directly solve \cref{euler-lagrange} because the derivatives of the machine learning model had too much noise. In Ref. \cite{snyder2012finding} noise was reduced by projecting functional derivatives onto a subspace spanned by relevant vectors via principal component analysis. A similar approach was taken in Ref. \cite{meyer2020machine}, where they projected the functional derivatives onto a subspace spanned by a sinusoidal basis. Here, without any projection scheme, we show that it is possible to use \cref{opt} to solve for an electron density directly. A projection scheme is not necessary since no derivatives are being taken with respect to the DNN. The VDNN directly outputs the kinetic energy and the functional derivative of the kinetic energy. We used a value of $\alpha=10^{-3}$ and a uniform electron density as the starting guess. We re-normalized the electron density at every step to enforce charge conservation and deemed a calculation converged when the absolute change of the energy between subsequent steps was less than $10^{-4}$ Ha.  The exact electron density and the electron density found using the VDNN are shown in \cref{fig:TF}. The densities differ minimally, and the total energy difference found between the two calculations was 19.1 meV / electron. Thus, machine learning models can be used in direct minimization calculations for the Thomas-Fermi model.  \\

\begin{figure}[t]
    \centering
    \includegraphics[width=\linewidth]{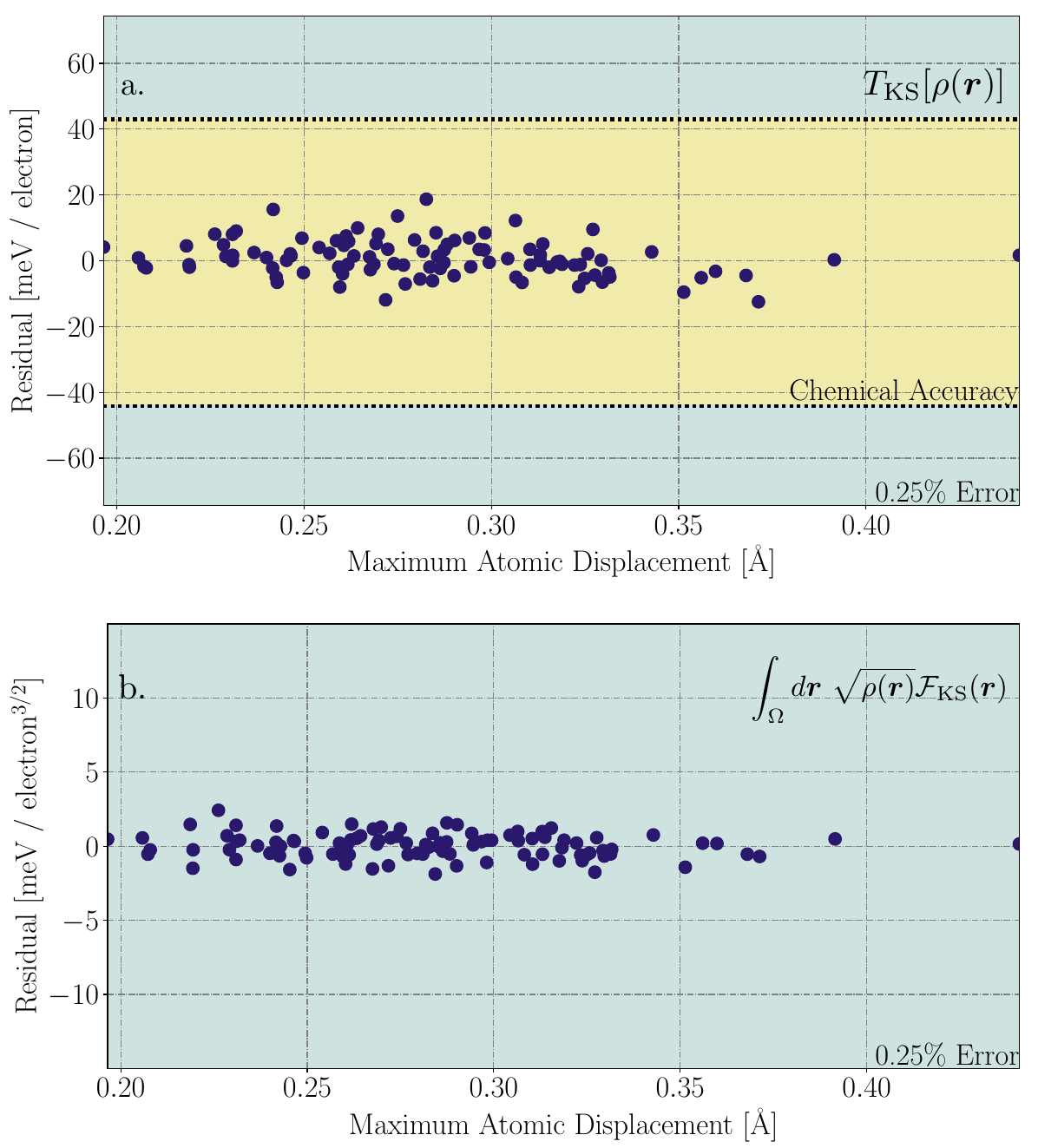}
    \caption{Residuals of (a) $T_{\text{KS}}[\rho(\bm{r})]$ and (b) $\int_{\Omega} d\bm{r}~\sqrt{\rho(\bm{r})}\mathcal{F}_{\text{KS}}(\bm{r})$ versus maximum atomic displacement for the Kohn-Sham model. Predictions are for a test set containing 100 graphene systems with 32 atoms as described in Section \ref{methods}. VDNNs allow for accurate predictions from small Kohn-Sham density functional theory datasets.}
    \label{fig:results_2}
\end{figure}

We now consider $\mathcal{T}_{\text{KS}}$ and $\mathcal{F}_{\text{KS}}$. After generating a training dataset that uniformly sampled $\sqrt{\rho}\mathcal{F}_{\text{KS}}$, we  trained a VDNN on $\mathcal{T}_{\text{KS}}$ and $\sqrt{\rho}\mathcal{F}_{\text{KS}}$ simultaneously with $\rho$, $\partial_x \rho$, and $\partial_y \rho$ as inputs. We found that including gradients as input channels reduced the mean squared error on the validation set by 7\%. Training on $\sqrt{\rho}\mathcal{F}_{\text{KS}}$ rather than $\mathcal{F}_{\text{KS}}$ reduced the mean squared error on the validation set by 43\%. Multiplication of $\sqrt{\rho}$ eliminates $\mathcal{F}_{\text{KS}}$ where $\rho=0$, and enhances $\mathcal{F}_{\text{KS}}$ where $\rho\neq0$. This filter-like behaviour allows for an improvement in the predictions. In \cref{fig:results_2}a, we plot the maximum atomic displacement versus residual energy per electron for the 100 testing atomic configurations. To determine percentage errors, the mean of the true kinetic energy values was used. From here, we see that all predictions are within chemical accuracy (43.4 meV). The MAE and RMSE were 4.3 meV / electron and 5.6 meV / electron respectively.  In \cref{fig:results_2}b, we plot true versus residual values for $\int_{\Omega} d\bm{r}~\sqrt{\rho(\bm{r})}\mathcal{F}_{\text{KS}}(\bm{r})$. From here we find that all values are well within 0.25\% error. MAE and the RMSE were 0.67 meV / electron${}^{3/2}$ and 0.83 meV / electron${}^{3/2}$.  VDNNs can provide all of the relevant information needed in OF-DFT. In addition, we find there is no increase in error as the maximum atomic displacement increases. Within the range of maximum atomic displacements, the error remains constant. However, using \cref{opt}, we were unable to obtain the correct electron density for the Kohn-Sham models. We also investigated bulk Al using a 4-atom unit cell with lattice constant of $a=2.856$ \AA. We followed the previous methodological protocol while only changing the k-point grid ($8\times8\times8$ grid). We also found for this system that we could not obtain the correct electron density via direct minimization with VDNNs. These failures, however, are not due to errors of the model, but due to the fact that we are only sampling $\mathcal{F}_{\text{KS}}$ for converged electron densities. In previous work \cite{snyder2012finding, meyer2020machine}, and for the KS-DFT data, functional derivatives of the kinetic energy are collected for only converged calculations. When using \cref{opt}, one encounters unconverged electron densities, and must also know the mapping from these unconverged electron densities to their respective kinetic energy densities and functional derivatives of the kinetic energies. Although $\mathcal{T}_{\text{KS}}$ is known for all iterations in a KS-DFT calculation, $\mathcal{F}_{\text{KS}}$ is not. The lack of samples of $\mathcal{F}_{\text{KS}}$ along the optimization path prevents the insertion of more accurate, kinetic energy machine learning frameworks in OF-DFT. This highlights the need for future work in this area. Solving this challenge would significantly reduce the amount of computation for accurate electronic structure calculations. It should be noted that unconverged densities found along an optimization path are also converged densities of another external potential. If one is able to access these external potentials, the sampling problem would be solved. A promising direction could be to use the differential virial theorem as demonstrated in Ref. \cite{ryabinkin2013exact}. As shown recently in Ref. \cite{ghasemi2021artificial}, this problem could also be solved by considering a differential equation that includes $\mathcal{F}$ and a source function. This source function depends explicitly on the electron density, and $\mathcal{F}$ can be found once this source function is known. Unfortunately, for KS-DFT calculations this source function is also only known for converged electron densities, but further work in this area could be promising.    \\

 We also investigated how VDNNs perform on a toy, 2 electron system in 1D previously investigated in Refs. \cite{snyder2012finding, mayer2020machine}. We used the same ResNet architecture and dataset as described in Ref. \cite{meyer2020machine}, with a field of view of 257 voxels for the VDNN and we compare our results to the ResNet model of Ref. \cite{meyer2020machine}. We found that using $\sqrt{\rho}\mathcal{F}$ also yielded a smaller mean squared error during training, as described above for $\mathcal{F}_{\text{KS}}$. For $T$,  our error was $\approx75$ times larger than Ref. \cite{meyer2020machine} with a MAE of 0.17 eV (3.84 kcal / mol). This large discrepancy is due to the previous models being trained directly on the energy rather than energy density. This allowed for highly accurate models with errors an order of magnitude less than chemical accuracy. For $\mathcal{F}$, we found that our error was $\approx1.9$ times larger with a MAE of 0.50 eV / electron (11.42 kcal / mol / electron). However, for $\mathcal{F}$, our maximum absolute error was 1.8 times smaller. In addition, when comparing the errors between $T$ for the 1D system and the 3D system ($T_{\text{KS}}$) we find an increase in error by a factor of $\approx20$ for the 1D system. As we change the number of physical dimensions, the number of inputs to the model increases. The number of pixels in the 3D case ($19^3$) is $\approx20$ times larger compared than the 1D case (257). In 3D, the network has more information to extract features from, which leads to more accurate predictions. It should also be noted that the VDNN is capable of performing inference for an arbitrarily sized 1D system, so long as the potentials and electron densities are similar to the training set. The existing models from Refs. \cite{snyder2012finding, mayer2020machine} are limited to the same system sizes used during training.\\ 

\begin{figure}[t]
    \centering
    \includegraphics[width=\linewidth]{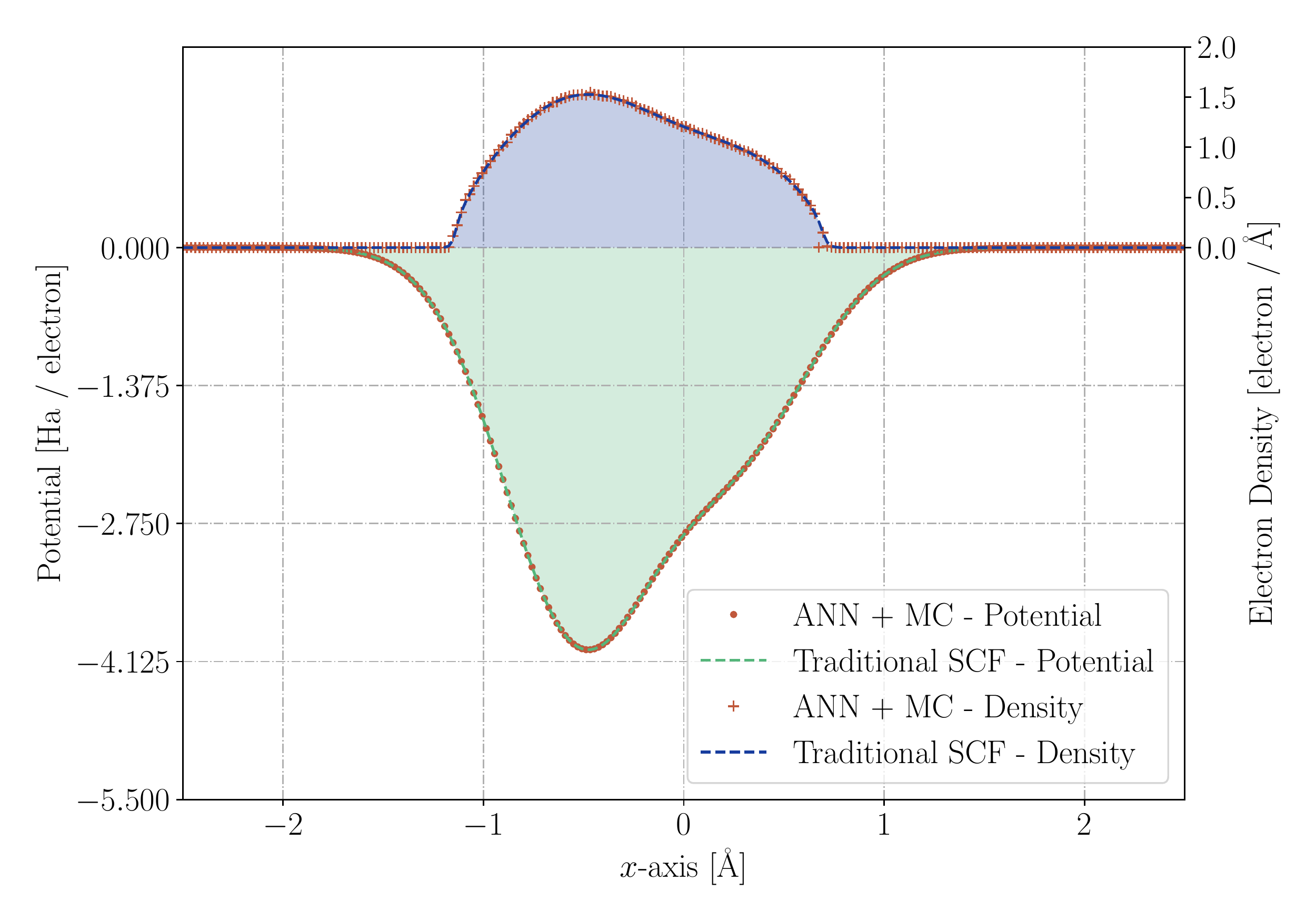}
    \caption{Comparison of electron densities and potentials between a Monte Carlo optimization (red crosses and points) and a traditional self-consistent field calculation (blue and green dashed lines) for 2 electrons in 1D with the external potential described in \cref{1d_energy}. The Monte Carlo optimization yields a very similar electron density to a self-consistent field calculation.}
    \label{fig_5}
\end{figure}

An alternative approach to minimizing \cref{total_energy} that avoids computing functional derivatives is MC optimization via the Metropolis algorithm \cite{metropolis1953equation}. Direct minimization approaches often require information about derivatives to make a gradient based update. Gradient free optimization is an alternative approach that does not require such information and is more compatible  with machine learning methods since the computational cost associated with inference is low and derivatives can be unreliable. To showcase this potential solution, we consider 2 electrons in 1 dimension with the Thomas-Fermi model as the kinetic energy functional. Using this approximation allows us to compare our MC based optimization with a traditional, gradient based optimization. The total energy functional, excluding exchange-correlation effects, can be written as
\begin{eqnarray}
    \label{1d_energy}
    E[\rho] &=& \frac{\pi^2}{12}\int_{\ell}dx~\rho^3(x) + \frac{1}{2}\int_{\ell}dx~\int_{\ell}dx'~\frac{\rho(x')\rho(x)}{|x - x'|} \nonumber\\
    &+&\int_{\ell}dx~\sum_{i=1}^2 -\alpha_i\exp(-(x - \beta_i)^2)\rho(x)
\end{eqnarray}
where $\ell$ is the length of the 1 dimensional cell. In \cref{1d_energy}, the first term is the kinetic energy of the 1 dimensional Thomas-Fermi model, the second term is the 1 dimensional Hartree energy, and the third term is the external energy from a toy, double inverted Gaussian potential. For the external energy, we used the parameters: $\alpha_1=1.0$ Ha / electron, $\alpha_1=2.0$ Ha / electron, $\beta_1=-0.5$ Bohr, $\beta_2=1.0$ Bohr. For the kinetic energy, we trained a 3 layer, fully connected neural network that maps a value of $\rho$ to a value of the one dimensional kinetic energy density. We generated $10^5$ random numbers from 0 to 1, which represented values of density, and trained the network for 100 epochs with a batch size of 100 and a learning rate of $10^{-5}$. We did not perform any standardization or normalization and we used ELU activation functions throughout the network. For the MC simulation, we performed simulated annealing with a starting value of $\beta^{-1}=10^{-4}$ Ha which was decreased according to the formula $\beta_{\text{new}}^{-1} = \beta_{\text{old}}^{-1} / (1.0 + 2\times10^{-6})^n$, where $n$ is the iteration number. After 2 million iterations, $\beta^{-1}=1.87\times10^{-6}$ Ha. At each iteration, we updated all values of $\rho$ in two steps. The first step was computing a random change
\begin{equation}
    \Delta\rho = 1000 (\rho(x) + 10\sigma)  u(\sigma, x) 
\end{equation}
where $\sigma$ is the standard deviation of a normal distribution and $u(\sigma, x)$ is function generated from a normal distribution centered at zero with the same shape as $\rho(x)$. The random change is then updated according to
\begin{equation}
    \Delta\rho = \Delta\rho - \rho\frac{\langle\Delta \rho\rangle}{\langle\rho\rangle},
\end{equation}
where $\langle f \rangle$ is the mean of $f$.
We used a standard deviation of $\sigma=10^{-5}$ which was reduced during the simulation following the same protocol as $\beta$. All proposed values of $\rho$ that were negative were set to zero, and $\rho$ was re-normalized at every step before evaluating the energy. In \cref{fig_5}, we plot $\rho$ and the potential (Hartree + external) for a traditional direct minimization calculation, following \cref{opt} alongside $\rho$ and the potential for the MC simulation. For the traditional gradient based calculation, the energy was declared converged when the difference in energy between subsequent steps was $<10^{-6}$ Ha. There is excellent agreement between the MC optimization and the gradient based optimization. The mean absolute differences between the charge densities, potentials, and total energies were $3.74\times10^{-3}$ electron / \AA, $5.61\times10^{-5}$ meV / electron, and 1.70 meV respectively. Future work involves implementing a 3 dimensional, functional derivative-free, OF-MC algorithm capable of calculating more accurate electron densities with improved, machine-learned kinetic energy functionals.

\section{Conclusion}
\label{conclusion}
We have shown that VDNNs can be used to accurately predict the kinetic energy density and the functional derivative of the kinetic energy for Kohn-Sham and Thomas-Fermi theories. This methodology drastically reduces the number of electronic structure calculations needed to generate a training set. We have shown that one can obtain an accurate charge density and total energy after training with data from only 1 direct minimization calculation for the Thomas-Fermi model. Similarly, we have shown that we can calculate accurate kinetic energies from only 2 converged calculations for Kohn-Sham density functional theory. Additionally, we show that this accuracy is held to arbitrary system size. Currently, one cannot use voxel deep neural networks to solve for Kohn-Sham electron densities via direct minimization. This is due to the fact that one only obtains the functional derivative of the kinetic energy from a Kohn-Sham calculation when convergence has been reached. However, unconverged electron densities found along an optimization path are also converged electron densities with another external potential. If these external potentials are found, the functional derivative of the kinetic energy would be known along an optimization path and one could use a voxel deep neural network in a direct minimization calculation. In addition, we show that an alternative, functional derivative-free, Monte Carlo based orbital-free algorithm could also be used to determine ground state electron densities.

\section{Acknowledgements}
The authors acknowledge fruitful discussion with Pierre Darancet and for comments on the manuscript. The authors also acknowledge Compute Canada, the National Research Council, and the Vector Institute for Artificial Intelligence for computational resources. KR acknowledges the National Sciences and Engineering Council of Canada, and the Vector Institute for Artificial Intelligence for funding. Work at the National Research Council was carried out under the auspicious of the AI4D program.

\section{Supplemental Information}
The supplemental information (SI) provides some hyperparameter convergence results as well as other results that accompanies the main text.

\subsection{Hyperparameter Studies and Additional Results}

\begin{figure}[t]
    \centering
    \includegraphics[width=0.8\linewidth]{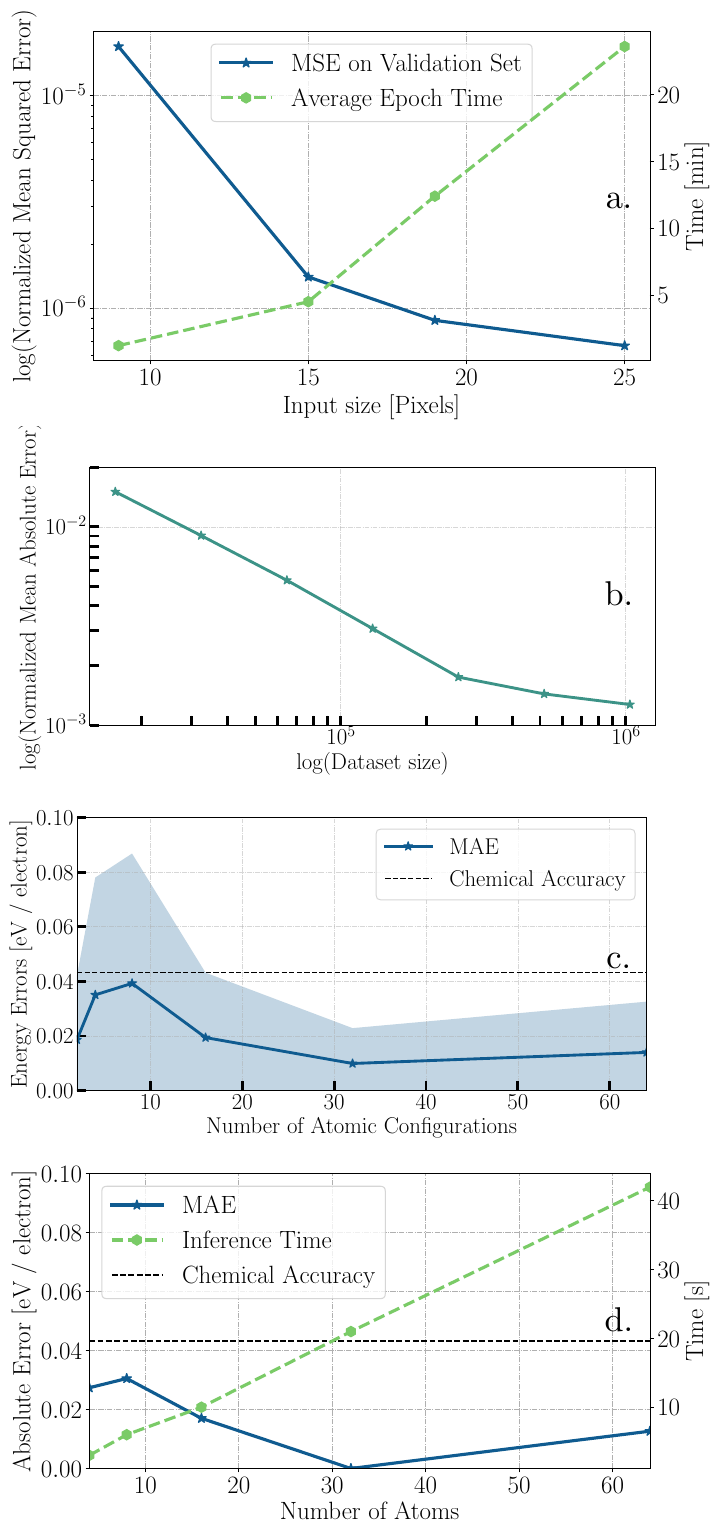}
    \caption{Convergence results for voxel deep neural networks. (a) The normalized mean squared error and the epoch time versus input size. (b) The normalized mean absolute error as a function of training dataset size. (c) The mean absolute error (line and points), and the root mean squared error (shaded region) as a function of the number of DFT training calculations. (d) The absolute error of the kinetic energy as a function of the number of atoms as well as the inference time versus the number of atoms. All kinetic energies shown here are from the Kohn-Sham non-interacting kinetic energy functional.}
    \label{fig:results_1}
\end{figure}

Before using VDNNs in practice, we focus on determining hyper-parameters using $\mathcal{T}_{\text{KS}}$.\\

To answer the question of optimal input size, we trained VDNNs with different input sizes and compared errors of different models. In \cref{fig:results_1}a, we show the normalized mean squared error of the validation sets as a function of input size. The length of inputs in each dimension is the same. For example, the input size of 19 corresponds to an input image with dimensions $19^3$. From \cref{fig:results_1}a, we see that as the image size is increased, the error decreases. We also see that the error is converging;  beyond a certain input size, the addition of extra pixels is not advantageous. As we increase the input size, the training and inference computational cost also increase. This can also be seen in  \cref{fig:results_1}a, where we plot the average epoch time as a function of input size. In this case, the computational cost increases linearly with the number of pixels. Thus when one chooses an input size, there is a balance between accuracy and computational cost. We found input sizes of $19^3$ were a good trade-off between accuracy and computational cost. \\

How many input examples are needed to produce an accurate model? To answer this question, we trained VDNNs with different training set sizes and compared the normalized mean absolute errors of the validation sets. In \cref{fig:results_1}b, we plot the normalized mean absolute errors of the validation sets as a function of training dataset size. From these plots, it is clear that the normalized mean absolute error converges as a function of the dataset size, and is well converged with a dataset size of $10^6$ images. This value was used when training all reported models unless otherwise stated. We note that a single SCF step produces $n_x \times n_y \times n_z$ samples, where $n$ denotes the number of real space grid points in a given direction. For the 32 atom graphene lattice, this number was $1.728\times10^6$. A single DFT calculation thus generates a large number of training examples and therefore very few DFT calculations are needed. We also see the slope of the line change at a dataset size of $\approx2.5\times10^5$ indicating a decrease in the rate of convergence. Based on this, one should use a minimum of $2.5\times10^5$ training examples to decrease the training time while maintaining accuracy. Again, this data can be easily extracted from DFT calculations.\\

How many calculations are needed to produce accurate kinetic energies? To answer this question, we trained VDNNs on the KS-DFT data and studied the accuracy of the models as a function of the number of atomic configurations. Specifically, we extracted a training dataset from 2, 4, 8, 16, 32, and 64 different training atomic configurations and calculated the mean absolute errors (MAEs), and root mean squared errors (RMSEs) of the kinetic energies for the testing set. It should be noted that a shift was applied to the predictions from VDNNs to obtain better results after integration. In a machine learning model, errors are never eliminated and become non-negligible after integrating on large numerical grids. A rigid shift on the \textit{training} set rids the error accumulation on both the training and testing sets. In  \cref{fig:results_1}c, we plot the MSE with their respective standard deviations. From the plot, we notice that error does not substantially decrease as a function of the number of atomic configurations. We, therefore, conclude that a model could be made from a training dataset with only 2 atomic configurations given that the MSE is less than chemical accuracy. Only 2 DFT calculations are needed to produce an accurate KED for pristine graphene lattices. \\

One of the major advantages of VDNNs is that they scale to arbitrary system size. After training a VDNN on the KS-DFT data, we ran calculations with the same kinetic energy cutoff (45 Ha) for 4, 8, 16, 32, and 64 atom unit cells. In \cref{fig:results_1}b, we show the absolute error of the predicted kinetic energy per electron and the inference time as a function of the number of atoms. From here, we see that the error remains constant as the number of atoms increases. In theory, VDNNs scale to an arbitrary system size with no increase in error per electron. The cost of inference scales linearly with the number of atoms (or number of grid points) in the system. The timings of the inference calculations were done with 16 nodes, each with 4 NVIDIA V100 GPUs.\\


\bibliography{refs}

\begin{figure*}[ht]
    \centering
    \includegraphics[width=\linewidth]{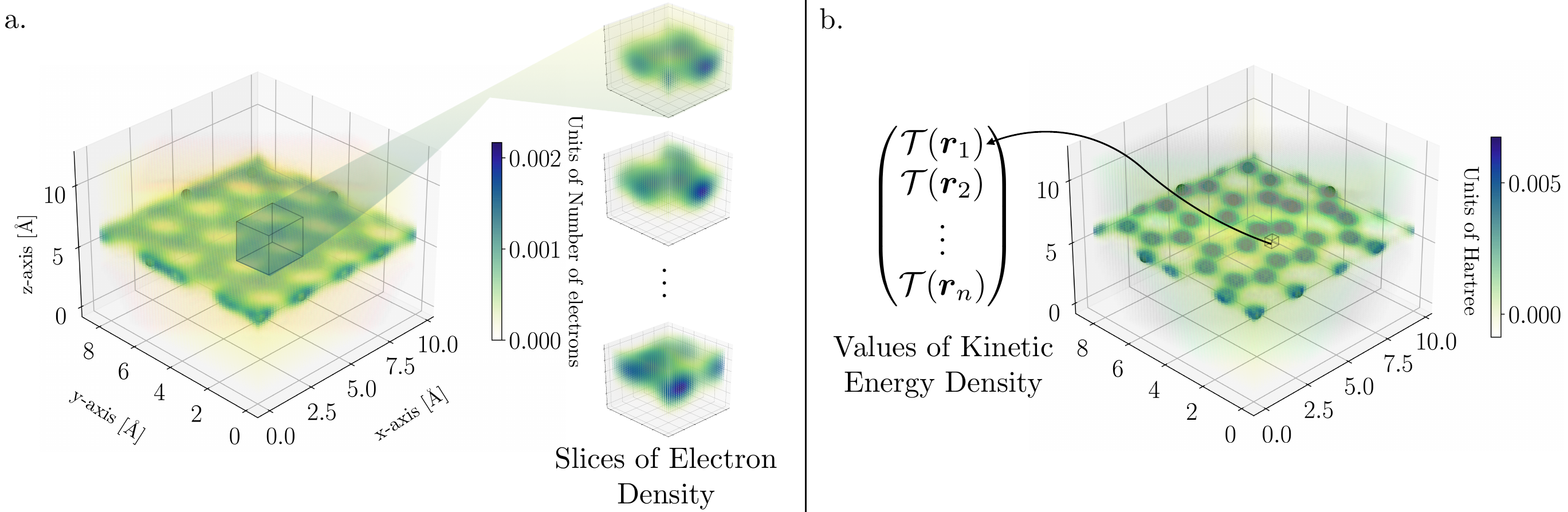}
    \caption{TOC Graphic}
\end{figure*}

\end{document}